\documentclass[a4paper]{jpconf}
\usepackage{graphicx}

\usepackage[normalem]{ulem}  
\usepackage[dvips]{color} 
\newcommand{\com}[1]{{\sf\color[rgb]{0,0,1}{#1}}}

\renewcommand\sout{\bgroup \color{red} \ULdepth=-.5ex \ULset}

\begin{document}
\title{(Anti-)strangeness production in heavy-ion collisions}

\author{P. Moreau$^{1,2}$,  F. Li$^{3}$, C.-M. Ko$^{3}$, W. Cassing$^4$,  E. L. Bratkovskaya$^{1,2}$}

\address{$^1$ Frankfurt Institute for Advanced Studies, Johann Wolfgang Goethe Universit\"at,
 Frankfurt am Main, Germany}
\address{$^2$ Institut f\"ur Theoretische Physik, Johann Wolfgang Goethe Universit\"at,
	Frankfurt am Main, Germany}
\address{$^3$ Cyclotron Institute and Department of Physics and Astronomy, 
	Texas A\&M University, College Station,
USA}
\address{$^4$ Institut f\"ur Theoretische Physik, Universit\"at Gie\ss{}en,
Germany}

\ead{moreau@fias.uni-frankfurt.de}

\begin{abstract} The production and dynamics of strange and antistrange hadrons
in heavy-ion reactions from $\sqrt{s_{NN}} \approx$ 3 GeV to 200 GeV is analyzed within the Parton-Hadron-String-Dynamics (PHSD) transport model. The PHSD results for strange baryon and antibaryon production are roughly consistent with the experimental data starting from upper SPS energies. Nevertheless, hadronic final state flavor-exchange reactions are important for the actual abundances, in particular at large rapidities where hadronic dynamics, parton fragmentation and string decay dominate. A striking disagreement between the PHSD results and the available data persists, however, for bombarding energies below  $\sqrt{s_{NN}} \approx$  8 GeV where the strangeness production is significantly underestimated as in earlier HSD studies. This finding implies that the strangeness enhancement seen experimentally at FAIR/NICA energies cannot be attributed to a deconfinement phase transition or crossover but probably involves the approximate restoration of chiral symmetry in the hadronic phase.
\end{abstract}

The production of strange hadrons in heavy-ion collisions has long been
suggested to provide a sensible probe for the heavy-ion dynamics and
its degrees-of-freedom at relativistic energies \cite{KochM,VPS10}. However, while the production of kaons and
antikaons in Pb+Pb collisions (158 A GeV or $\sqrt{s_{NN}}$ = 17.3
GeV) at top SPS energies turned out to be well described by
hadron/string models such as the Hadron-String-Dynamics (HSD)  
or URQMD \cite{Br04}, the
abundances of antistrange baryons were clearly underestimated. Furthermore, the kaon and hyperon abundances have been
underestimated by up to a factor of two at the lower bombarding
energies of 4 to 40 A GeV \cite{Br04}. This surprising result has led to the suggestion that a quark-gluon plasma (QGP) might  
have already been produced at  lower energies \cite{yy}.  But this raises
 the question  why we did not see a similar strangeness enhancement at the
top super-proton-synchrotron (SPS) or even the Relativistic Heavy Ion Collider (RHIC)
energies where a QGP is more likely produced.
The answer to this problem is not straightforward since to
differentiate between the hadronic and partonic degrees-of-freedom
and their impacts on the various hadronic final-state interactions
requires a fully microscopic description of these reactions. To
this aim, an off-shell covariant transport approach -- denoted as the Parton-Hadron-String Dynamics (PHSD) -- has been developed to incorporate all necessary ingredients. This model  has proven in the past to provide a good description of light hadron abundances, collective
flow and electromagnetic radiation from nucleus-nucleus collisions
\cite{Volodya2014,PHSD,PHSDrhic,To12,KBC12,Linnyk2011} up to
the Large-Hadron-Collider (LHC) energies.

We recall that the PHSD model is a covariant dynamical approach for
strongly interacting systems formulated on the basis of
Kadanoff-Baym equations~\cite{JCG04,Ca09} or off-shell transport
equations in the phase-space representation. In the
Kadanoff-Baym theory, the field quanta are described in terms of
dressed propagators with complex selfenergies. Whereas the real part
of the selfenergies can be related to mean-field potentials of
Lorentz scalar or vector type, the imaginary parts provide
information about the lifetime and/or reaction rates of time-like
particles~\cite{Ca09}. Once the proper complex selfenergies of the
degrees-of-freedom are known, the time evolution of the system is
fully governed by off-shell transport equations for quarks and
hadrons (as described in Refs.~\cite{JCG04,Ca09}). The PHSD model
includes the creation of massive quarks via hadronic string decay to partons 
above the critical energy density $\sim 0.5$ GeV/fm$^3$ and quark
fusion forming a hadron in the hadronization process. With some
caution, the latter process can be considered as a simulation of a
crossover transition since the underlying EoS in PHSD is a
crossover~\cite{Ca09}. At energy densities close to the critical
energy density, the PHSD describes a coexistence of the quark-hadron
mixture. This approach allows for a simple and transparent
interpretation of lattice QCD results for thermodynamic quantities
as well as correlators and leads to effective strongly interacting
partonic quasiparticles with broad spectral functions. For a review
on off-shell transport theory we refer the reader to
Ref.~\cite{Ca09}. Results from the PHSD model and their 
comparison with experimental observables for heavy-ion collisions from the lower
SPS  to RHIC energies can be found in
Refs.~\cite{Volodya2014,To12,KBC12,Linnyk2011}. In the hadronic
phase, i.e. for energies densities below the critical energy
density, the PHSD approach is identical to the
HSD model~\cite{HSD,HSD2}, which has been
successfully employed for p-p, p-A and A-A reactions up to top SPS
energies.

We here present an overview of (anti-) strangeness production in
central Au+Au (Pb+Pb) collisions from $\sqrt{s_{NN}} \approx$ 3 GeV
to 200 GeV within the PHSD with a particular focus on the
multi-antistrange baryon production. Compared to the HSD study in Ref.
\cite{HSD2}, we now have incorporated 
in the hadronic sector all
strangeness exchange reactions $\pi+N \leftrightarrow K+Y, N+N
\leftrightarrow N+K+Y, {\bar K}+N \leftrightarrow \pi+Y$,  $\pi+ \Xi
\leftrightarrow {\bar K}+Y,  Y+Y \leftrightarrow N+\Xi$, and ${\bar
K}+\Xi \leftrightarrow \Omega + \pi,  \Xi + \Xi \leftrightarrow
\Omega + Y$ in meson-baryon and baryon-baryon collisions following
Refs. \cite{Ko,Ko1,Ko2}. It is essential to recall that in PHSD (as
in HSD) the dominant three-body channels $B+{\bar B} \leftrightarrow
$3 mesons are included on the basis of 'detailed-balance'
\cite{Ca02}, where the dominant mesons involve the $\pi, \rho, \omega,
K, K^*$ degrees-of-freedom.

\begin{figure}[h]
 	\includegraphics[width=0.49\textwidth]{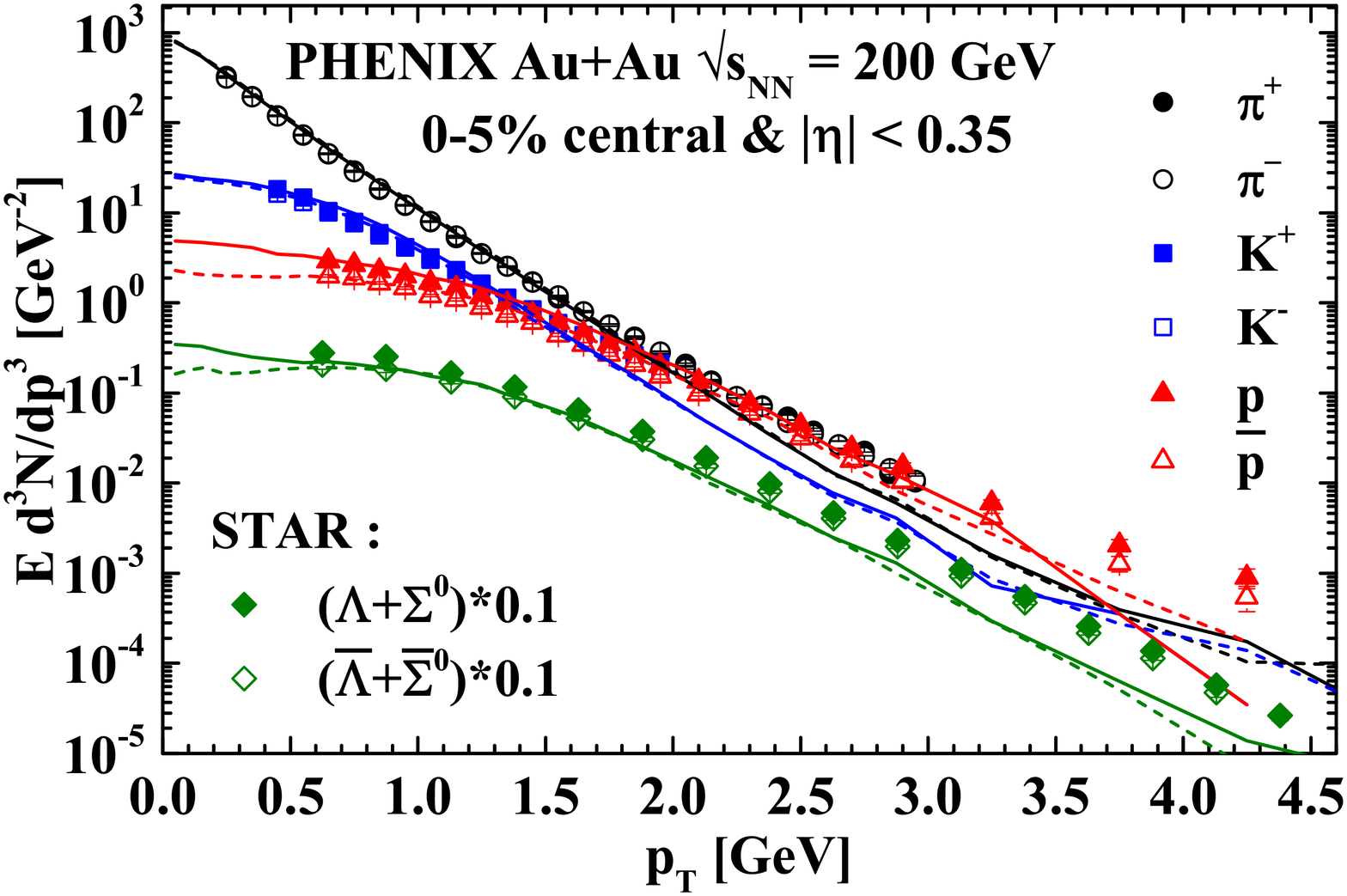}	\includegraphics[width=0.48\textwidth]{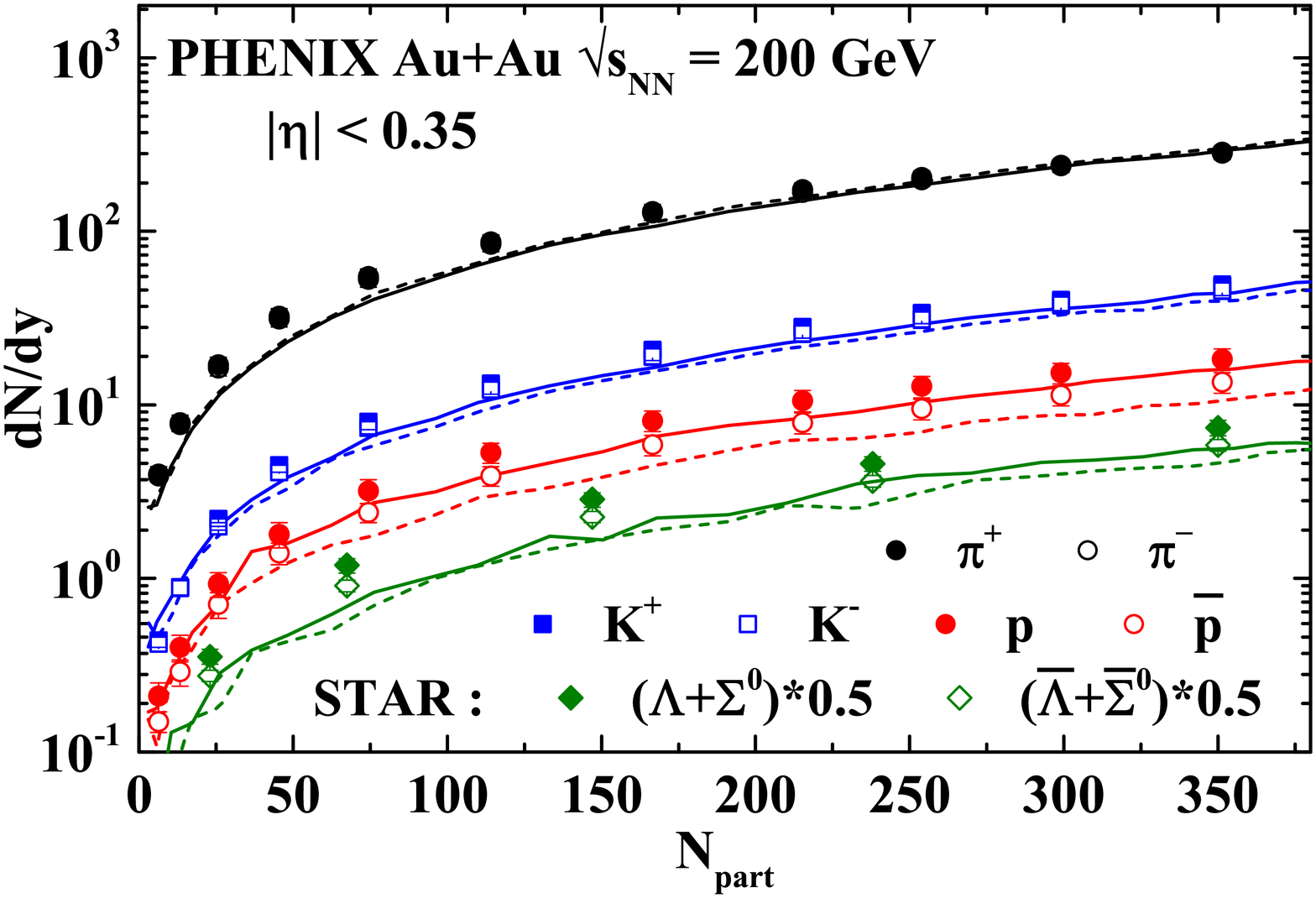}
 	\caption{The invariant yield $Ed^3N/dp^3$ as a function of the transverse momentum $p_T$ (left window) and 
the yield $	dN/dy$ as a function of the number of participating nucleons $N_{part}$ (right window) at midrapidity for pions, kaons, protons, antiprotons, $\Lambda$ and $\bar{\Lambda}$ for Au+Au collisions at $\sqrt{s_{NN}}=200$ GeV in comparison to the experimental data from the PHENIX \cite{PHENIX} and STAR collaboration \cite{STAR}.}
 	\label{fig1}
\end{figure}
We start with PHSD results for the transverse momentum distributions
of particles (solid lines) and antiparticles (dashed lines), shown in the left window of Fig.
1 for 5\% central Au+Au collisions at $\sqrt{s_{NN}}$ = 200
GeV in comparison to the data from PHENIX and STAR
\cite{PHENIX,STAR}. Apart from a reasonable description of the data,
we find baryons and antibaryons to differ only at low $p_T < $ 0.5
GeV/c where some  net absorptions of antibaryons becomes visible.
The centrality dependence of the particle yield (at
midrapidity) is displayed in the right window of Fig. 1 as a function of the
number of participants $N_{part}$ in comparison to the PHENIX and
STAR data \cite{PHENIX,STAR}. Here we find only a small difference
between protons and antiprotons, whereas hyperons and
antihyperons are about the same roughly in line with the data. We
mention that the differences between baryons and antibaryons
increase drastically when going down in bombarding energy, where
nucleons from the target and projectile are shifted to the
midrapidity region and dominate over antibaryons, which are
preferentially produced in the hadronization process from the QGP at midrapidity.
\begin{figure}[h]
	\includegraphics[width=0.49\textwidth]{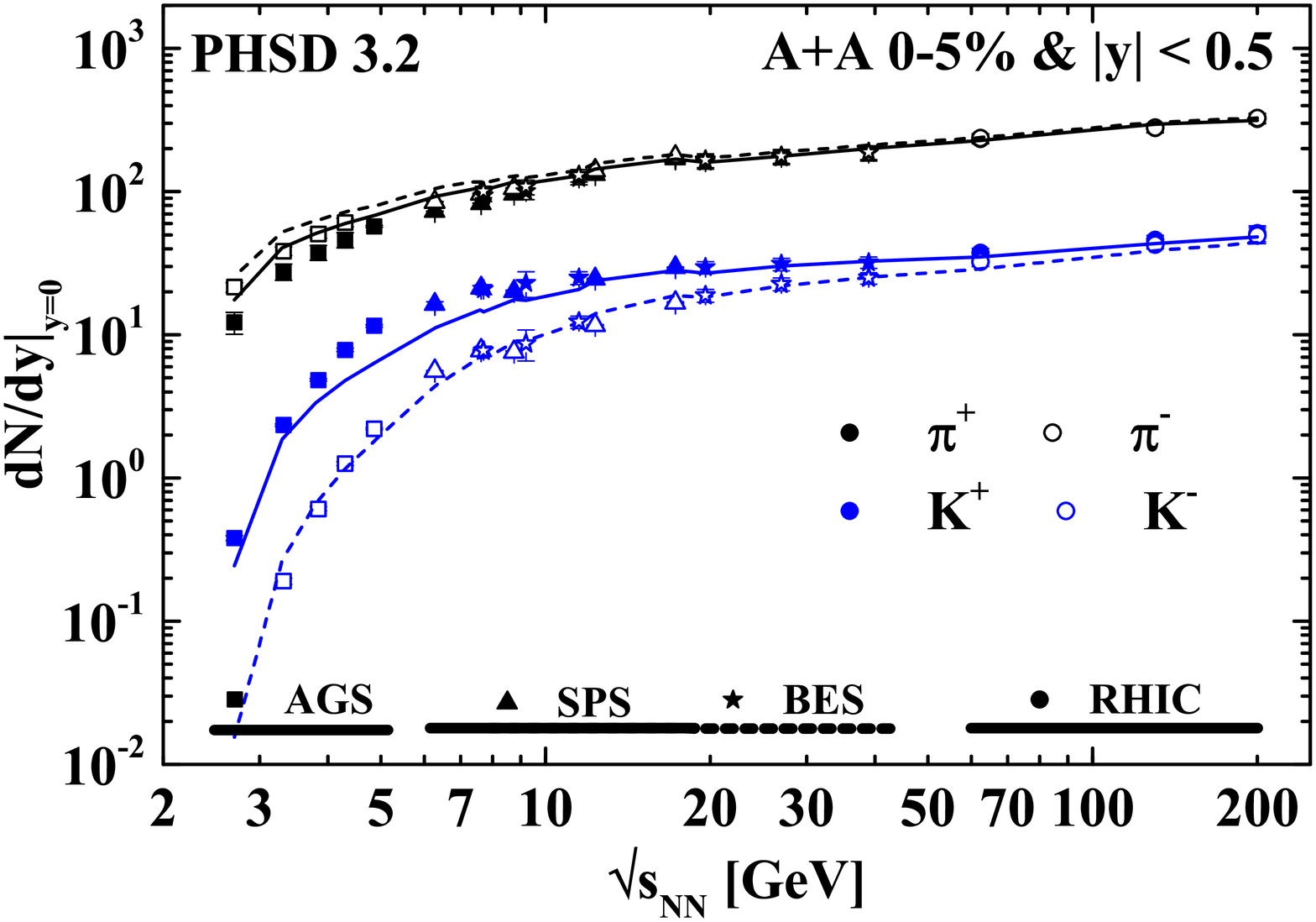}	\includegraphics[width=0.49\textwidth]{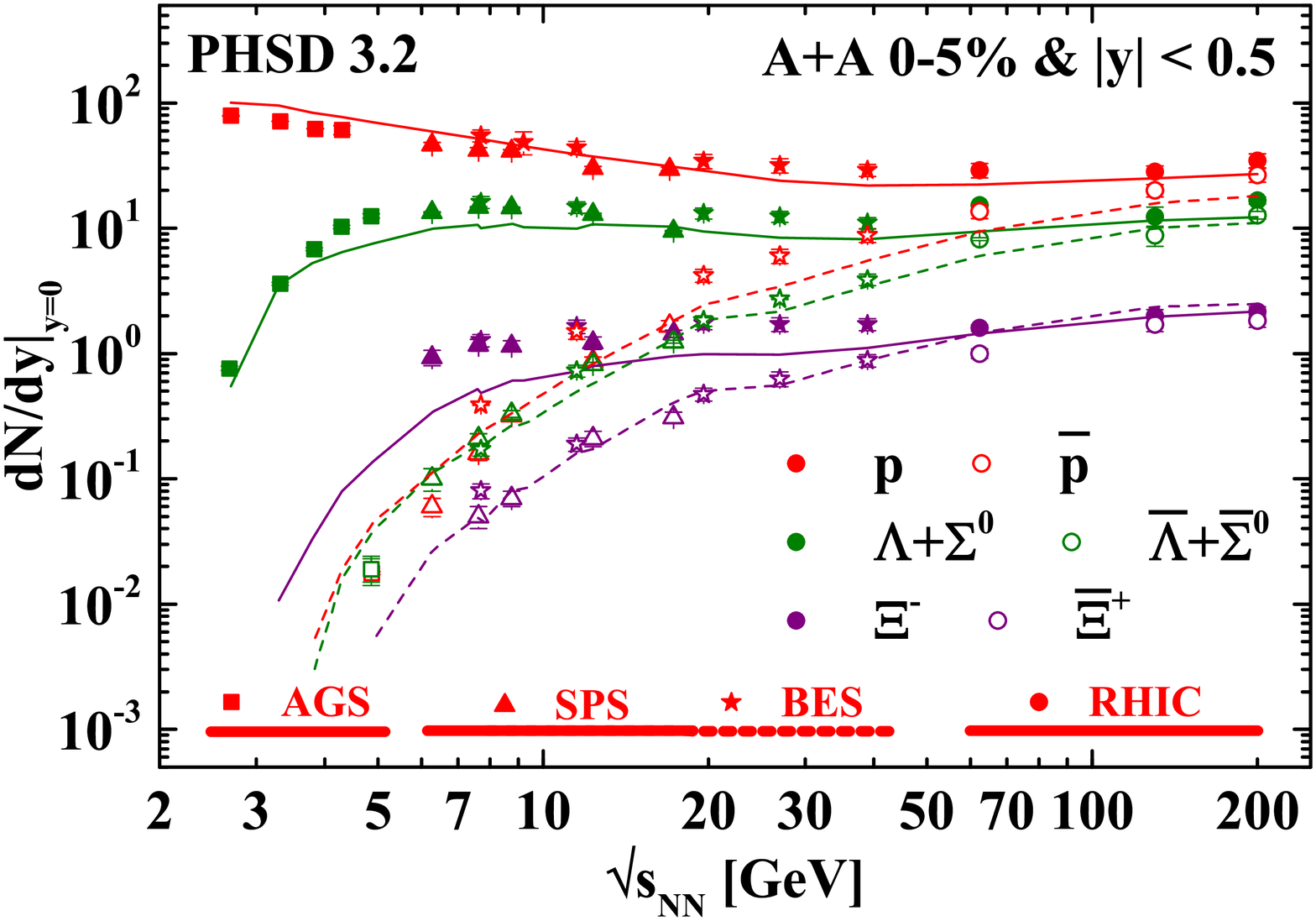}
	\caption{(\protect Left window) The yield at midrapidity for pions and kaons as a function of $\sqrt{s_{NN}}$ for central Au+Au and Pb+Pb collisions in comparison to the experimental data from \cite{AGS_mesons} for AGS energies, from \cite{SPS_mesons} for SPS energies, from \cite{BES} for the preliminary results from the BES program at RHIC, and from \cite{TOP_RHIC} for RHIC energies. (\protect Right window) The yield at midrapidity for protons, antiprotons, $\Lambda$, $\bar{\Lambda}$, $\Xi^-$ and $\bar{\Xi}^+$ as a function of $\sqrt{s_{NN}}$ for central Au+Au and Pb+Pb collisions in comparison to the experimental data from \cite{AGS_baryons} for AGS energies, from \cite{SPS_baryons} for SPS energies, from \cite{BES} for the preliminary results from the BES program at RHIC, and from \cite{TOP_RHIC,RHIC_strange_baryons} for RHIC energies.}
	\label{fig2}
\end{figure}

We further display in the left window of Fig. 2 the excitation function for
$\pi^\pm$ and $K^\pm$ mesons from 5\% central Au+Au collisions (at
midrapidity) as a function of $\sqrt{s_{NN}}$ in comparison to the
available data. We observe an overestimation of pion
production by PHSD at low invariant energy and an underestimation of
$K^+$ mesons as in the HSD model \cite{Br04}. An underestimation
of strangeness at low invariant energy is also seen in the
baryon-antibaryon sector (cf. Fig. 2 (right window)) where the hyperons ($\Lambda + \Sigma^0$)
and cascade baryons ($\Xi^-$)  are missed substantially. On the other hand, the antihyperons
(${\bar \Lambda} + {\bar \Sigma^0}$) and $\Xi^+$ are reasonably described, and this
can be traced back to the
dominant production in the hadronization process from the QGP.

A closer analysis of the explicit rapidity distribution of
baryons and antibaryons shows that antibaryon production 
close to midrapidity can essentially be attributed to hadronization from the 
QGP\com{,} whereas
at forward or backward rapidities (closer to the target and
projectile rapidities) hadrons are mainly produced from fragmentation or
hadronic string decay in the PHSD. This allows for a scan in the
partonic fraction of the dynamics when analyzing separately baryons
and antibaryons as a function of rapidity.

In summary:  we have extended the PHSD transport approach to include all strangeness exchange reactions $\pi+N \leftrightarrow
K+Y, N+N \leftrightarrow N+K+Y, {\bar K}+N \leftrightarrow \pi+Y$,
$\pi+ \Xi \leftrightarrow {\bar K}+Y,  Y+Y \leftrightarrow N+\Xi$,
and ${\bar K}+\Xi \leftrightarrow \Omega + \pi,  \Xi + \Xi
\leftrightarrow \Omega + Y$ in meson-baryon and baryon-baryon
collisions\com{,} thus allowing for a detailed study of multistrange baryon
production in the hadronic phase and to compare 
their production in the hadronization process
from the QGP. We find that an underestimation of strangeness
production at low invariant energy is seen in the meson as well as
baryon sector (cf. Fig. 2), whereas the multistrange
baryon/antibaryon  dynamics at the top SPS and RHIC energy are rather well
described by hadronization from the QGP at midrapidity. This finding
implies that the strangeness enhancement seen experimentally at
FAIR/NICA (AGS and lower SPS) energies cannot be attributed to a
deconfinement phase transition (or crossover) but probably involves the approximate restoration of chiral symmetry in the hadronic phase, which 
would lead to a change of hadron properties and thus an enhanced production of multistrange baryons.
\vspace{-0.2cm}
\section*{Acknowledgements}
The authors acknowledge the support by BMBF, HIC for FAIR and the HGS-HIRe for FAIR.
The computational resources were provided by the LOEWE-CSC.

\end{document}